\documentstyle[twocolumn,aps,epsf,floats]{revtex}
\begin{document}
{\bf Comment on: ``Using Ni Substitution and $^{17}$O NMR to Probe the 
Susceptibility $\chi'({\bf q})$ in Cuprates''}
\vskip 0.2cm

In a recent letter, Bobroff {\em et al.}~\cite{BAY97}
presented  novel $^{17}$O NMR measurements for
YBa$_2$(Cu$_{1-x}$Ni$_x$)$_3$O$_{6+y}$. 
They observed a strong $T$-dependent broadening of the $^{17}$O 
NMR-lines which they attributed to the oscillatory electron spin polarization 
induced by Ni impurities. 
Their experiment offers a new probe of the momentum
dependence of the static spin susceptibility $\chi'({\bf q})$,
complementary to the NMR observation of the Gaussian 
component of the transverse
relaxation time, $T_{\rm 2G}$, of planar Cu~\cite{PS91}.

To understand the strong $T$
dependence of the NMR  line width $\Delta \nu(T)$,
Bobroff {\em et al.}~performed calculations to simulate the 
NMR line shape by assuming a Gaussian form for the electron spin 
susceptibility 
$\chi_{\rm G}'({\bf q})=4\pi\chi^* \xi^2 \exp[-({\bf q}-{\bf Q})^2 \xi^2]$ 
with ${\bf Q}=(\pi,\pi)$.  
They found that the $^{17}$O line width $\Delta \nu$ is 
independent of the antiferromagnetic correlation length $\xi$. For the 
overdoped sample ($y=1$), where $\Delta \nu=\chi^* f(\xi)$ is only very weakly 
$T$-dependent, they conclude that $\chi^*$ is basically $T$-independent. 
However, for the underdoped samples ($y=0.6$) they find that the strong 
$T$-dependence of $\Delta \nu$
can {\it only} be  explained with a $T$-dependent $\chi^*$.
Combining these results with the $T$-dependence of $T_{2G}^{-1} \sim 
\chi^* \xi$, they pointed out that this implies a $T$-independent $\xi$ for 
the underdoped samples. They also remark that a Lorentzian model  
$\chi_{\rm L}'({\bf q})=4\pi\chi^* \xi^2/(1+({\bf q}-{\bf Q})^2 \xi^2)$
gives similar results.
This is in contradiction to the spin fluctuation scenario of cuprate 
superconductors~\cite{BP95}, which is based on the 
Lorentzian form $\chi_{\rm L}'({\bf q})$.  

Stimulated by their experiment we also performed calculations to simulate 
the $^{17}$O NMR lineshape. 
For the Gaussian susceptibility $\chi_{\rm G}'({\bf q})$, we obtain 
the same results as Bobroff {\em et al.}
However, we obtain a strong $\xi$ dependence of the $^{17}$O 
line width with the Lorentzian form of $\chi_{\rm L}'({\bf q})$.
Our  results using $\chi_{\rm L}'({\bf q})$
are shown in Fig.~1, where we plot the $\xi$-dependence 
of $\Delta \nu$.
Due to the $1/T$ dependence of the Ni magnetic moment, $\Delta \nu$ 
corresponds to $T \Delta \nu_{\rm imp}$ in Ref.~\cite{BAY97}.
The inset shows our results for $\chi_{\rm G}'({\bf q})$.
Our results obtained with $\chi_{\rm L}'({\bf q})$ (curve {\it (a)}) 
demonstrate that the experimental results of 
Bobroff {\em et al.} are clearly compatible with a $T$-dependent
$\xi$~\cite{BP95}.
Furthermore, including in addition to the nearest neighbor Cu-O hyperfine 
coupling $C$ a next-nearest neighbor coupling $C^\prime$ \cite{Zha} 
(curves {\it (b), (c)} and {\it (d)}) we obtain a flattening of $\Delta \nu(\xi)$ 
for $\xi=1..2$.
This provides a possible explanation for the different behavior of 
overdoped ($\xi=1..2$) and underdoped ($\xi=2..4$) systems.\\
\begin{figure} [t]
\begin{center}
\leavevmode
\epsfxsize=7.5cm
\epsffile{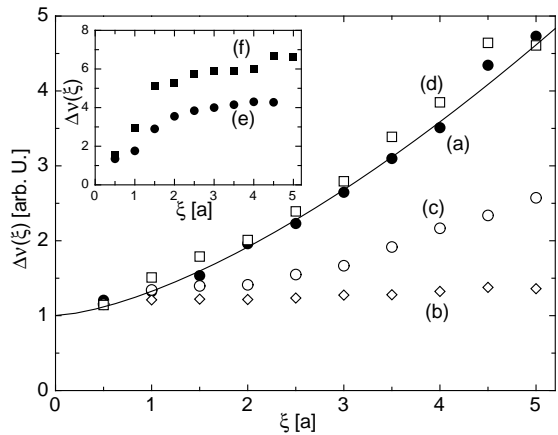}
\end{center}
\caption{The $^{17}$O linewidth $\Delta \nu$ as a function of $\xi$. 
Curve {\it (a)} shows the result for  $x=2 \%$ Ni doping and $C^\prime=0$. 
The solid line is a fit with $\Delta \nu=1.0+0.32\xi^{3/2}$. 
Curves {\it (b), (c)} and $\it (d)$ correspond to $x=0.5, 2$ and 
$4 \%$ Ni doping, respectively and $C^\prime/C=0.25$.  The inset 
represents the results for  $\chi_{\rm G}'({\bf q})$ with $ (e)$ $x=2 \%$ 
and $(f)$ $x=4 \%$ Ni doping.}
\end{figure} 
Due of  the location of the $^{17}$O between two $^{63}$Cu sites, the local field at the $^{17}$O site behaves as $\sim \partial \tilde{\chi}(r)/\partial r$, where $\tilde{\chi}(r)$ is the envelope of the real space susceptibility (see Fig.~4 in Ref.~\cite{BAY97}). Our analytical computations show that for the Gaussian form $\chi_{\rm G}'({\bf q})$, $\Delta \nu(\xi)$ is approximately constant for a realistic range of $\xi$. For the Lorentzian form $\chi_{\rm L}'({\bf q})$, $\Delta \nu(\xi) \sim \xi^{3/2}$, in agreement with our numerical 
results (see solid line in Fig.~1).

Taking the $T_{2G}$ data from \cite{Tak} (corrected for $T_1$ contributions \cite{Curro}) and $T \Delta \nu$ for the underdoped sample from \cite{BAY97}, we computed the product $T_{2G} T \Delta \nu$, which is independent of $\chi^*$, and {\it which for any form of $\chi^\prime (q)$ } depends solely on $\xi$. In contrast to \cite{BAY97} we 
find that this product is strongly $T$-dependent, dropping by more than a factor of 2 between $100 \, K$ and $200 \, K$. For a Gaussian this implies that $\xi$ increases as $T$ increases, an unreasonable result. For a Lorentzian, $\xi$ decreases with increasing $T$. We therefore believe that a Gaussian 
form of $\chi^\prime (q)$ is unlikely.\\
This work has been supported by STCS under NSF Grant No.~DMR91-20000, 
the U.S. DOE Division of Materials Research under 
Grant No.~DEFG02-91ER45439 (C.P.S., R.S.) and the Deutsche 
Forschungsgemeinschaft (J.S.)

\vskip 0.3cm
\noindent
D. K. Morr, J. Schmalian, R. Stern, and C.P. Slichter\\
University of Illinois at Urbana-Champaign\\
Loomis Laboratory of Physics\\
1110 W. Green, Urbana, Il, 61801


\begin{thebibliography}{99}
\bibitem{BAY97} J. Bobroff {\em et al.} Phys. Rev. Lett. {\bf 79}, 2117 (1997).
\bibitem{PS91} C.H. Pennington and C.P. Slichter, Phys. Rev. Lett. {\bf 66}, 
381 (1991). 
\bibitem{BP95}  D. Pines, Z. Phys. B {\bf 103},  129  
(1997), and references therein.
\bibitem{Zha} Y. Zha, V. Barzykin and D. Pines, Phys. Rev. B {\bf 54}, 7561 
(1996).
\bibitem{Tak} M.~Takigawa, Phys. Rev. B {\bf 49}, 4158 
(1994).
\bibitem{Curro} N. Curro {\it et al.}, Phys. Rev. B {\bf 56}, 877 
(1997).
\end{thebibliography}
\end{document}